\documentclass[%
reprint, 
superscriptaddress,
 amsmath,amssymb,
 aps,
 longbibliography,
]{revtex4-1}
\usepackage{hyperref}
\usepackage{cleveref}
\usepackage{graphicx}
\usepackage{dcolumn}
\usepackage{bm}

\usepackage{braket}
\usepackage{letltxmacro}

\LetLtxMacro{\ORIGselectlanguage}{\selectlanguage}
\makeatletter
\DeclareRobustCommand{\selectlanguage}[1]{%
  \@ifundefined{alias@\string#1}
    {\ORIGselectlanguage{#1}}
    {\begingroup\edef\x{\endgroup
       \noexpand\ORIGselectlanguage{\@nameuse{alias@#1}}}\x}%
}
\newcommand{\definelanguagealias}[2]{%
  \@namedef{alias@#1}{#2}%
}
\makeatother
\definelanguagealias{en}{english}
\definelanguagealias{En}{english}
\definelanguagealias{EN}{english}
\definelanguagealias{{EN}}{english}

\begin{document}

\preprint{APS/123-QED}

\title{Delayed transition to coherent emission in nanolasers with extended gain media
}

\author{F. Lohof}
\affiliation{Institute for Theoretical Physics, Otto-Hahn-Allee 1, P.O. Box 330440, University of Bremen}
\author{R. Barzel}
\affiliation{Institute for Theoretical Physics, Otto-Hahn-Allee 1, P.O. Box 330440, University of Bremen}
\author{P. Gartner}
\affiliation{Centre International de Formation et de Recherche Avanc\'ees en 
Physique - National Institute of Materials Physics, P.O. Box MG-7, 
Bucharest-M\u{a}gurele, Romania}
\author{C. Gies}
\affiliation{Institute for Theoretical Physics, Otto-Hahn-Allee 1, P.O. Box 330440, University of Bremen}

\date{\today}

\begin{abstract}
The realization of high-$\beta$ lasers is one of the prime applications of
cavity-QED promising ultra-low thresholds, integrability and reduced power
consumption in the field of \textit{green photonics}. In such nanolasers spontaneous
emission can play a central role even above the threshold. By going beyond rate-equation approaches, we
revisit the definition of a laser threshold in terms of the input-output
characteristics and the degree of coherence of the emission. We demonstrate
that there are new regimes of cavity-QED lasing, realized e.g. in high-$Q$
nanolasers with extended gain material, for which the two can differ
significantly such that coherence is reached at much higher pump powers than
required to observe the thresholdlike intensity jump. Against the common perception,
such devices do not benefit from high-$\beta$ factors in terms of power
reduction, as a significant amount of stimulated emission is required to quieten
the spontaneous emission noise.
\end{abstract}

\pacs{Valid PACS appear here}
\maketitle


\section{Introduction}
Lasers have been around for almost 60 years and most of the underlying physics is
well established and understood. The threshold, in particular, is the central defining property of any laser device, as it separates the regime of spontaneous emission from that of phase-coherent operation. In conventional laser devices, the threshold can easily be identified from the input-output characteristics alone: Below threshold, a majority of photons spontaneously emitted from the gain material
is lost, and only a fraction described by the $\beta$ factor ends up in the
modes that can ultimately achieve lasing. Above threshold, stimulated emission
into these modes completely overpowers the losses, which gives rise to an
intensity jump typically over several orders of magnitude, which clearly marks the
laser threshold (``intensity threshold''). This fundamental behavior can
already be understood in terms of coupled rate equations for the excitation of
the gain material and the photon number \cite{yokoyama_rate_1989,
bjork_analysis_1991}.

In recent times, the laser threshold has been revisited due to the uprise of nanolasers that operate in new cavity-enhanced lasing regimes \cite{chow_emission_2014,rice_photon_1994,blood_laser_2013}.  \textit{Nature Photonics} was amongst the first journals to issue a laser ``checklist'' to ensure a certain standard in identifying laser operation, which becomes even more important for small lasers \cite{noauthor_scrutinizing_2017}. Cavities with
ultra-small mode volumes with dimensions of the light's wavelength allow for tinkering with
spontaneous emission itself. Quantum-electrodynamical effects make it possible
to enhance the emission rate \cite{wang_giant_2016,strauf_single_2011}, to suppress emission into nonlasing
modes \cite{stepanov_highly_2015, gevaux_enhancement_2006,
takiguchi_enhanced_2013}, or do both at the same time -- thereby fundamentally
changing the contributions from spontaneous and stimulated emission. When spontaneous emission is nearly completely directed into the laser mode
($\beta\approx1$), the intensity jump even disappears and makes an
identification of the threshold from the input-output curve alone impossible
 \cite{prieto_near_2015,
ota_thresholdless_2017,khajavikhan_thresholdless_2012}. In
such thresholdless lasers, it has become common practice to investigate fluctuations in the
photon number $n$ that are captured in the two-photon correlation function \cite{chow_emission_2014, strauf_self-tuned_2006, ulrich_photon_2007}
\begin{equation}
 g^{(2)}(0) = \frac{\langle n^2\rangle-\langle n\rangle}{\langle n\rangle^2}~.
 \label{eq:g2}
\end{equation}
For coherent emission, $g^ {(2)}(0)$ takes on the value of 1, whereas {for spontaneously emitted
light below threshold it} has a fingerprint $g^ {(2)}(0)=2$. As we show, the increased role of spontaneous emission in nanolasers can substantially delay the formation of coherence to higher pump powers.

In the present work, we refine the definition of the laser threshold by going beyond the rate-equation approximation to take a combined look at the intensity threshold and the transition of the
emission from thermal to coherent light reflected in $g^{(2)}(0)$
(``coherence threshold''). While one might intuitively believe that both
characteristics are interlinked and one necessitates the other, we show that
there are operational regimes of nanolasers that give rise to a separation between the two thresholds, so that the emission becomes coherent at much higher excitations, after the intensity
threshold has been crossed. An important result of this finding is that one
commonly used criterion for lasing in cavity-enhanced nanolasers, i.e.~the mean intracavity photon number $n
= 1$, can be arbitrarily insufficient and truly
only holds in the limit of the single-emitter laser \cite{nomura_laser_2010,
strauf_single_2011,gies_single_2011}. Furthermore, in those regimes an increase of the $\beta$ factor has no benefit in terms of
reducing the threshold power, an aspect that has, to the best of our knowledge,
not been recognized in the literature so far. The implications are both of
fundamental relevance for understanding the laser threshold, and apply to
present-day nanolaser devices. Despite the simplicity of our model, it captures all relevant effects of the laser dynamics and allows for the definition of analytical expressions for the above-mentioned threshold criteria.


\section{Characterizing the laser threshold beyond the rate-equation approximation}


At the rate-equation level the laser threshold is defined in terms of the
output intensity. This criterion we define as the intensity threshold. In conventional lasers it is well established that the threshold marking the transition to coherent emission is accompanied by a
sharp increase of output intensity. This intensity jump is less pronounced or
vanishes completely in high-$\beta$ lasers. Here we consider $N$ independent two-level emitters in resonant Jaynes-Cummings
interaction with a single mode of an optical cavity
 \cite{gartner_laser_2016}. This simplified laser model provides analytic access
to quantities related to the threshold, yet it contains the relevant physics to
render its implications applicable to current cavity-QED nanolaser devices.
Starting from the system Hamiltonian we derive equations of motion for the
upper-level population $f$ (the lower-level population is $1-f$) and the
intracavity photon number $n$ that resemble standard laser rate equations
 \cite{coldren_diode_2011}
(see Appendix A):
\begin{subequations}
\begin{align}
\dot{f}&=-R\left[n(2f-1)+f\right]-\gamma f+P(1-f),\label{RE1}\\
\dot{n}&=RN\left[n(2f-1)+f\right]-\kappa n,\label{RE2}
\end{align}\label{RE}
\end{subequations}
where
\begin{equation}
R=\frac{4|g|^2}{P+\gamma+\kappa}\label{Rc}
\end{equation}
is the spontaneous emission rate in the cavity mode, $g$ is the light-matter interaction strength, $\gamma$ is the rate of
radiative losses, $P$ is the pump rate, and $\kappa$ is the cavity decay rate.
The $\beta$ factor is defined as the ratio of the spontaneous emission into the
laser mode $R$ to the total spontaneous emission including losses, i.e.
\begin{equation}
\beta=R/(R+\gamma).\label{beta}
\end{equation}

In their stationary limit, the rate equations can be used to define the pump value $P_{\mathrm{th}}$ of the intensity threshold. In Appendix A, we show that it obeys the equation
\begin{equation}
P_\mathrm{th}-\gamma - (P_\mathrm{th}+\gamma)\frac{\kappa}{RN} = 0. \label{Pthr}
\end{equation}
\begin{figure}[t!]
   \centering
    \includegraphics[trim={2.5cm 0 2.5cm 0},clip,width=0.5\textwidth]{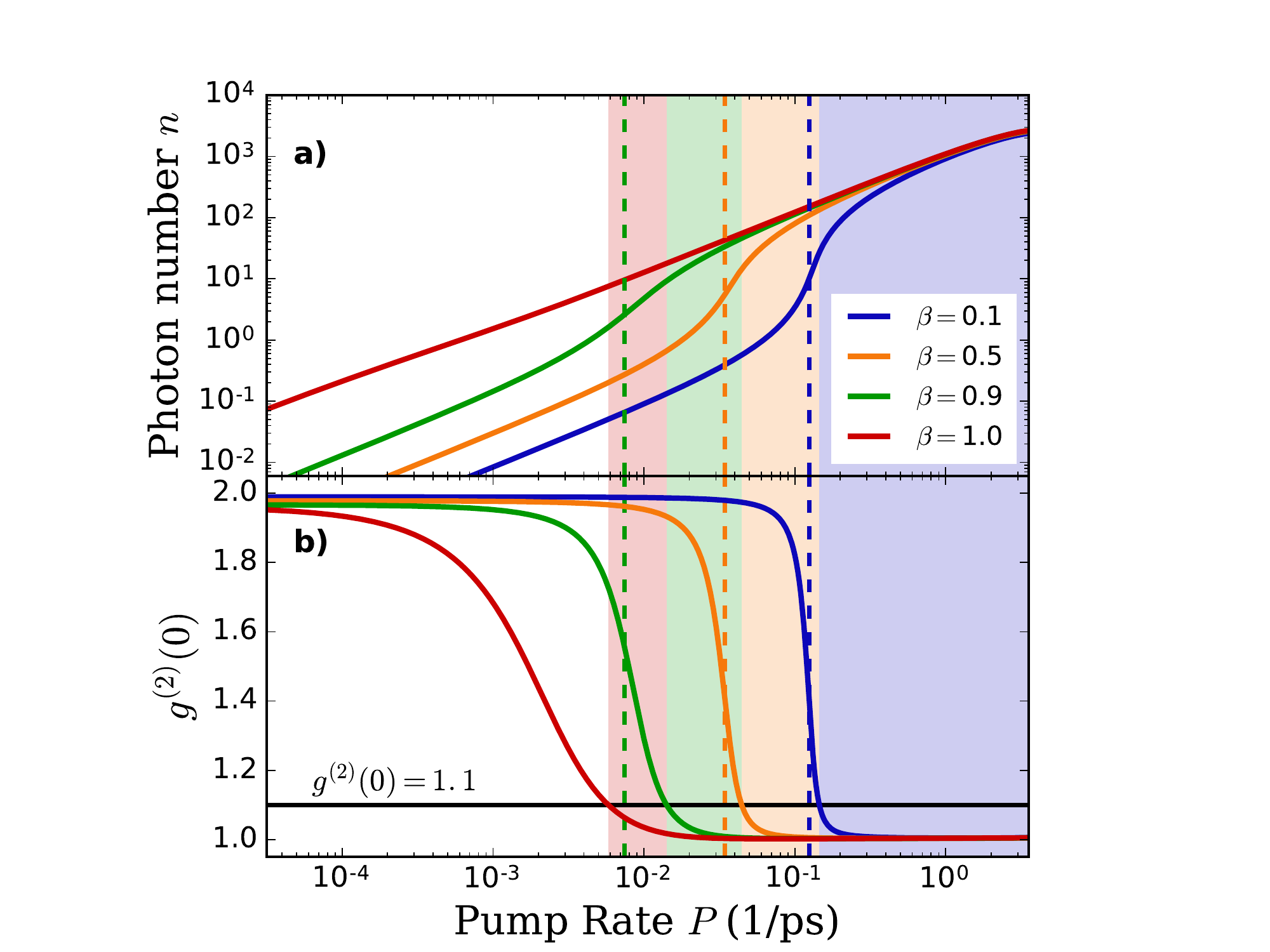}
    \caption{\textbf{a)} Output intensity and \textbf{b)} $g^{(2)}(0)$ values according to Eqs.~\eqref{RE} and \eqref{analyticg2} are shown for different values of $\beta$. The parameters $N=100$, $\kappa=0.04\,\mathrm{ps}^{-1}$, and $g=0.03\,\mathrm{ps}^{-1}$ correspond to typical high-$Q$ quantum dot (QD) nanolasers  \cite{ulrich_photon_2007}. The intensity thresholds are marked as dashed lines. In this parameter regime, they nearly coincide with the threshold to coherent emission, which are indicated by the shaded regions for each value of $\beta$. }\label{fig:Fig2}
 \end{figure}

The functionality of the threshold pump rate $P_{\mathrm{th}}$ defined this
way is illustrated in the top panel of Fig.~\ref{fig:Fig2}. For typical quantum-dot (QD) nanolaser parameters
taken from [\citenum{ulrich_photon_2007}], output intensities are shown (solid
lines) together with  the respective values for $P_{\mathrm{th}}$ marked by the
vertical lines. Different values of $\beta$ are used to characterize different
device efficiencies as explained below. As can be seen, the threshold pump
rates determined from Eq.~\eqref{Pthr} lie exactly in the middle of the
corresponding jump intervals in the output intensities.

Due to excitation-induced dephasing, the emission rate $R$ depends on the pump
rate. With it, $\beta$ is implicitly excitation dependent. Nevertheless, since
a constant $\beta$ factor is well established throughout the literature to
characterize laser efficiency, we choose to use $\beta$ rather than the
emission rate $\gamma$ into nonlasing modes, which is more commonly used in
theory. To do so, all given values of
$\beta$ are evaluated at $P=P_{\mathrm{th}}$. While nonconstant
$\beta$ factors have been investigated in different
contexts \cite{gericke_controlling_2018,
lichtmannecker_few-emitter_2017,gregersen_quantum-dot_2012}, this approximate
treatment fully serves the purpose in this work.

For $\beta$ approaching unity, the intensity jump goes to zero, reflecting the
concept of thresholdless lasing often referred to in the
literature \cite{prieto_near_2015, ota_thresholdless_2017,
khajavikhan_thresholdless_2012}. However, it is well established that coherent
emission can take place at finite excitation even in the limiting case
$\beta=1$ \cite{gies_semiconductor_2007}.  With no distinct threshold nonlinearity in the output intensity of
high-$\beta$ lasers, one has to rely on \emph{statistical} properties of the
emission in order to identify lasing \cite{strauf_self-tuned_2006,chow_emission_2014,ulrich_photon_2007}.
In a classical picture, the output intensity fluctuates around its mean value,
which allows distinguishing a thermal from a coherent light source.
The laser model we employ is fully quantum mechanical. In this picture, a
coherent (thermal) state of the light field is characterized by a Poissonian
(Bose-Einstein) photon distribution function $p_n$.
This quantity can be accessed both in
theory by using master-equation
approaches \cite{gartner_two-level_2011,jagsch_quantum_2018,
leymann_intensity_2013,carmele_analytical_2011}, and in experiment by using
photon-number-resolved detection
schemes \cite{schlottmann_exploring_2018,calkins_high_2013}.
In practice, it is much easier to use the second-order photon autocorrelation
function $g^{(2)}(0)$ to characterize the emission, as it can be readily
measured using Hanbury Brown and Twiss setups
 \cite{hanbury_brown_correlation_1956}. It is defined in Eq.~\eqref{eq:g2}, with $\langle n^k\rangle=\sum_n p_n n^k$, and the transition to its coherent value of 1 has become
an established method to characterize nanolasers that operate close to the
thresholdless regime with $\beta \approx
1$ \cite{strauf_self-tuned_2006,chow_emission_2014,ulrich_photon_2007}.

We access the two-photon correlation function beyond the rate-equation limit by extending Eqs.~\eqref{RE} to
contain higher-order correlations and truncate the arising hierarchy of quantum-mechanical expectation values up to and including two-photon correlations
 \cite{gies_semiconductor_2007}. This approach, the details of which are given in Appendix B, allows us to arrive at an analytic expression for the photon autocorrelation function that only
contains the photon number $n$, the carrier occupation $f$, and fixed system
parameters
\begin{equation}
g^{(2)}(0) =
2-\frac{\frac{R}{n}(2n+1)(n+f)+\frac{2\kappa}{N}}{\kappa+\frac{\Gamma}{2}
+R(2n+1)-\frac{R\Gamma}{2\kappa}N(2f-1)},\label{analyticg2}
\end{equation}
where we define $\Gamma=P+\gamma+\kappa$.
By using the stationary values for $f$ and $n$ provided
by the rate equations and taking advantage of the relationship between them
\begin{equation}
f = n\,\frac{NR+\kappa}{NR(2n+1)} \label{f_vs_n} \, ,
\end{equation}
which follows from Eq.~\eqref{RE2} in steady state, Eq.~\eqref{analyticg2} can be further simplified
to obtain a closed analytic expression for $g^{(2)}(0)$ in terms of the photon number only:
\begin{equation}
g^{(2)}(0)=2-\frac{R(2n+2)+\frac{3\kappa}{N}}{R(2n+1)+\kappa+\frac{\Gamma}{
2\kappa}\frac{NR+\kappa}{2n+1}}~.\label{g2_of_n}
\end{equation}
This expression is a key result of the paper, it provides access to the statistical properties of the emission even when values for the photon number obtained from rate equations are inserted. In Appendix B we show that the results obtained from Eq.~\eqref{g2_of_n} agree extremely well with full solution from the extended quadruplet equations so that, in fact, it can be used to augment results of rate-equation calculations with the corresponding $g^{(2)}(0)$ values.

In Fig.~\ref{fig:Fig2} the photon correlation function $g^{(2)}(0)$ as obtained
from Eq.~\eqref{analyticg2} is shown together with output intensities for
different $\beta$ factors. At low excitations $g^{(2)}(0)$ exhibits a value of 2. As it is customary in the literature, we model the excitation as an incoherent process, described by a Lindblad term (see Eq.~(A2)). It represents physically the scenario where intermediate scattering processes from higher excited ones to the final states before recombination dephase any phase information that may have been imprinted by the exciting laser. As a result, the emitted photons exhibit no phase coherence leading to thermal emission with $g^{(2)}(0)=2$. For all values of $\beta$, the transition of $g^{(2)}(0)$ to 1 (marked by the shaded regions) indicating coherent emission closely follows the intensity threshold (marked by the vertical lines). This behavior corresponds to the common conception of the intensity threshold being interlinked with the transition to coherent emission. Only for $\beta=1$ the intensity threshold vanishes and solely $g^{(2)}(0)$ indicates the existence of a transition.

We define the coherence threshold $P_\mathrm{coh}$ as the point, where
$g^{(2)}(0)$ indicates the onset of coherent emission. Here we choose $g^{(2)}(0)=1.1$ with the aim to identify the beginning of an interval in which $g^{(2)}(0)$ stays practically constant around unity. Of course, the choice of 1.1 for $g^{(2)}(0)$ as the onset of coherence is to a certain extent arbitrary, however, a lower value would only push the coherence and the intensity thresholds even further from one another. It is important to point out that a unique definition of $P_\mathrm{coh}$ is only
possible in the limit $\beta \rightarrow 0$. For large values of $\beta$ that
are relevant in nanolasers, $g^{(2)}(0)$ approaches 1 gradually over a wide range
of excitation powers.

 \begin{figure}[t!]
   \centering
    \includegraphics[trim={2.5cm 0 2.5cm 0},clip,width=0.5\textwidth]{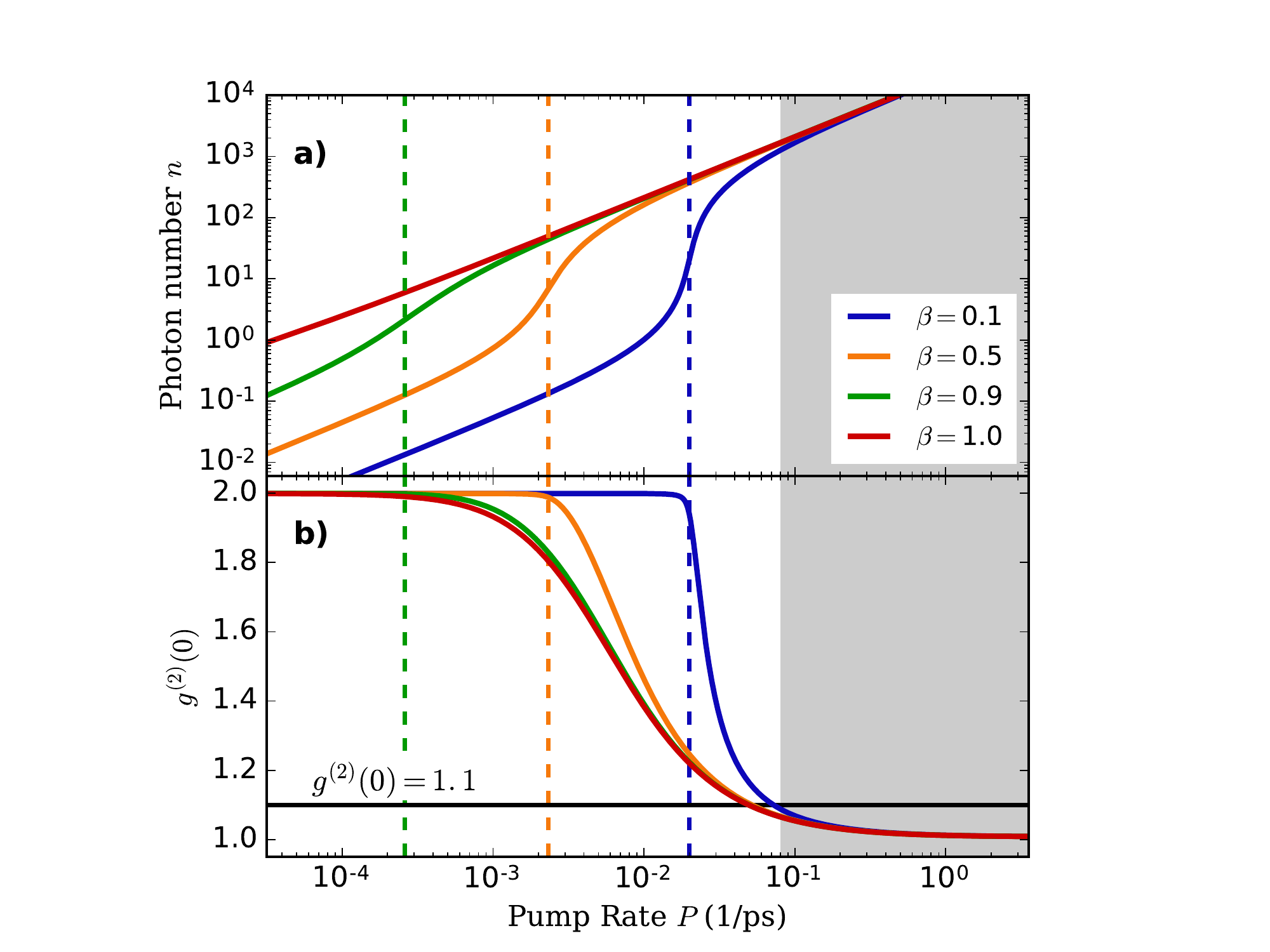}
    \caption{For a nanolaser with an extended gain material, \textbf{a)} output intensity and \textbf{b)} $g^{(2)}(0)$ values are shown for different $\beta$. The large carrier number in the gain media is reflected in $N=3\times 10^4$, whereas the other parameters are $\kappa=0.7\,\mathrm{ps}^{-1}$, $g=0.02\,\mathrm{ps}^{-1}$. For all values of $\beta$, the intensity threshold (dashed vertical lines) is clearly separated from the transition to coherent emission (indicated by the gray area). Furthermore, the coherence thresholds indicate that increasing $\beta$ has no benefit in terms of reducing the pump power to reach coherent emission.\label{fig:Fig3}}
 \end{figure}

\section{New Laser Regimes}
We now turn to a different class of nanolasers and show that they elude the common conception
of the laser threshold. Until now, high-$Q$ nanolasers were mostly realized in high-$Q$ cavities with discrete emitters, such as quantum dots, as gain
material. The previous section illustrated that for these systems, the
formation of coherence is linked to the intensity threshold as one typically
expects. A fundamentally different behavior is encountered in a very different,
yet highly relevant regime of cavity-QED nanolasers that operate with extended
gain materials instead of individual emitters. Atomically thin layers of
semiconducting transition-metal dichalcogenides (TMDs), such as WSe$_2$ or MoS$_2$,
have recently moved into the focus of optoelectronic device
applications \cite{tian_optoelectronic_2016} including nanolasers
 \cite{li_room-temperature_2017, salehzadeh_optically_2015, ye_monolayer_2015,
wu_monolayer_2015}. These new materials combine advantages in ease of
fabrication over conventional III/V semiconductors with a plethora of
possibilities to engineer their electronic and optical properties. The possibility of population inversion \cite{chernikov_population_2015} and room-temperature lasing with monolayer TMD flakes on top of photonic crystal \cite{wu_monolayer_2015} or nanobeam \cite{li_room-temperature_2017} cavities is being vividly discussed in the literature. Also recently, room-temperature lasing from a two-dimensional GaN quantum well (QW) embedded in a nanobeam cavity has been demonstrated \cite{jagsch_quantum_2018}.

Typical carrier densities in extended gain media (in the order of $10^{13}\,\mathrm{cm}^{-2}$ in
two-dimensional TMD materials) strongly exceed carrier numbers in discrete
gain media, such as quantum dots. This results in very different
operational regimes with severe implications regarding the threshold behavior.
While more complex semiconductor laser models can be employed
 \cite{jagsch_quantum_2018,hofmann_quantum_1999,chow_analysis_2015}, for the present purpose it suffices to bear with the
formalism used up to this point by mapping the number of electron-hole pairs in
the two-dimensional gain media to our discrete-emitter model. This notion is
formally justified, since the creation of cavity photons results from the
recombination of an electron-hole pair both in the case of $N$ independent
emitters each containing a single excitation, or $N$ excitations in an extended
system. Assuming an active region of $0.3\,\mathit{\mu m}^2$ and a
typical carrier density of $10^{13}\,\mathrm{cm}^{-2}$ yields an
estimate for the maximum number of excitons of $N=3\times 10^4$.
The parameters we use in
the following correspond to that of the nanobeam-quantum-well laser
investigated by Jagsch et al. \cite{jagsch_quantum_2018}.
In addition to much larger $N$, reported $Q$ factors for both the nanobeam and TMD nanolasers are
significantly lower and are typically around 2500
 \cite{jagsch_quantum_2018,li_room-temperature_2017, salehzadeh_optically_2015,
wu_monolayer_2015, ye_monolayer_2015}.

Input-output curves and photon-autocorrelation functions corresponding to the parameters of Ref.~[\citenum{jagsch_quantum_2018}], but also applicable to recent work on TMD gain
media, are shown in Fig.~\ref{fig:Fig3}. As for the few-emitter case
(Fig.~\ref{fig:Fig2}), the intensity threshold, marked by the intensity jump in
the input-output curves, moves to lower pump rates with increasing $\beta$
factor, suggesting that the threshold current can be strongly reduced by
maximizing the $\beta$ factor. The corresponding autocorrelation functions are
shown in the lower panel and reveal a very different picture: The transition to
coherent emission is strongly offset to higher pump rates from the intensity
jump, so that, in fact, the coherence and intensity thresholds are no longer
aligned. Coherent emission is reached at a pump rate of about $0.07/\mathrm{ps}$
\emph{irrespective} of the value of $\beta$, implying that for devices
with extended gain media, a reduction of the threshold current is no longer
possible by reducing radiative losses (i.e., increasing $\beta$).

\begin{figure}[t!]
   \centering
    \includegraphics[width=0.5\textwidth]{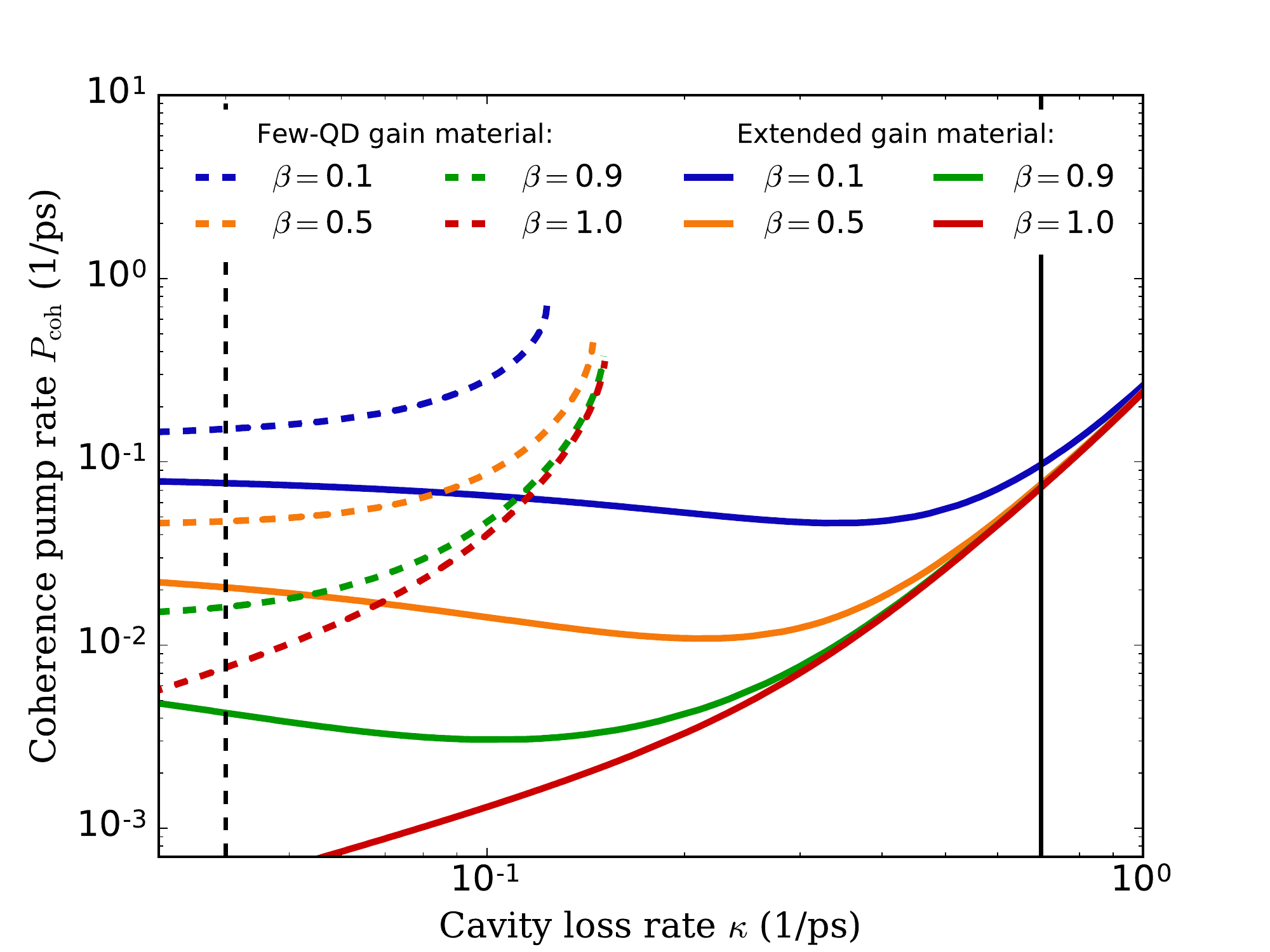}
    \caption{Pump rate $P_{\mathrm{coh}}$ at the coherence threshold as a function of cavity loss rate $\kappa$. Parameters ($N$, $g$) correspond to those in Figs.~\ref{fig:Fig2} (dashed) and \ref{fig:Fig3} (solid). Vertical lines indicate respective values for $\kappa$. One observes the different dependence on $\beta$ for the two cases as well as the transition between the regimes for varying $\kappa$. For small number of emitters there is an upper bound for $\kappa$ beyond which lasing can not be achieved and thus the dashed lines terminate above a certain value. \label{fig:Fig4}}
 \end{figure}

This behavior, which is a key insight of this work, can be attributed to the
very distinct parameter regimes: Room-temperature lasing from the extended gain
media in Refs.~[\citenum{jagsch_quantum_2018,li_room-temperature_2017}] takes
place in a regime, where cavity losses are the dominant quantity in determining
the laser threshold. Despite the relatively low $Q$ factors, lasing is still
possible due to the large number of carriers producing gain. For
few-emitter nanolasers, as the systems summarized in Ref.~[\citenum{strauf_single_2011}], a much larger
cavity $Q$ is required to attain lasing, which sets an upper bound for $\kappa$.
In this low-$\kappa$ regime, radiative losses dominate over the cavity losses,
so that the threshold pump rate is strongly dependent on $\beta$. These results
are summarized in Fig.~\ref{fig:Fig4}, which compares the pump rate $P_{coh}$,
at which coherent emission is reached, for the few-emitter (dashed lines) and
extended gain media (solid lines) as a function of the cavity loss rate. The cavity loss rates for both operational regimes we consider are indicated by vertical lines.

\begin{figure}[t!]
   \centering
    \includegraphics[width=0.5\textwidth]{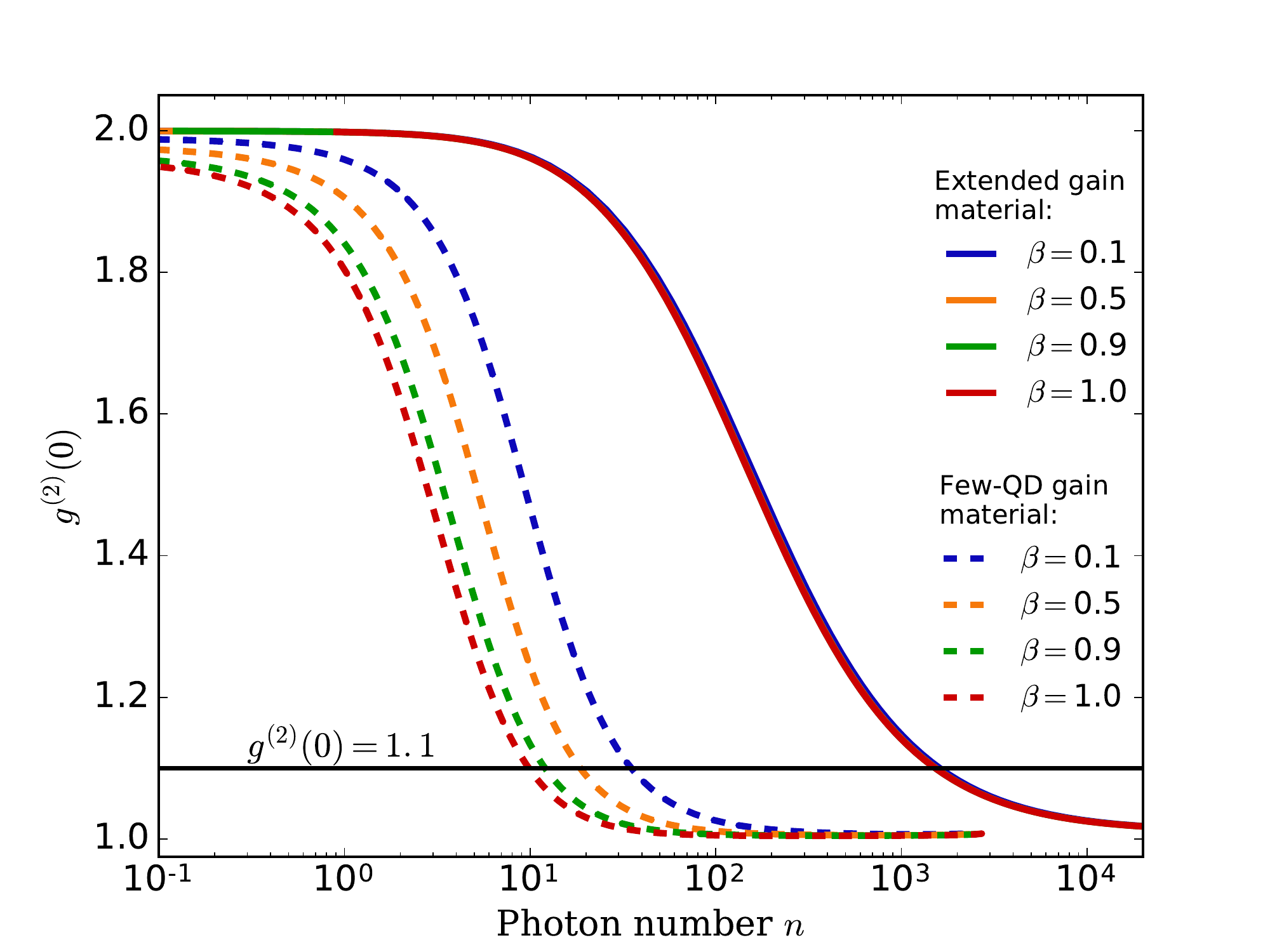}
    \caption{Two-photon correlation function $g^{(2)}(0)$ as a function of
 photon number. For high-$Q$ quantum-dot nanolasers (dashed lines) an increasing
 $\beta$ factor leads to fewer photons required for coherent emission. For a high number of emitters and lower cavity $Q$ (solid lines) coherent emission happens at much larger photon numbers and independent of $\beta$.\label{fig:Fig5}}
 \end{figure}

More insight is revealed by condensing the information in Fig.~\ref{fig:Fig3}
into a depiction of $g^{(2)}(0)$ as a function of the mean photon number.
Surprisingly, for the parameters of the extended gain media (solid lines),
Fig.~\ref{fig:Fig5} shows that the transition to coherent emission is solely
determined by the mean intracavity photon number completely irrespective of the
value of $\beta$. Coherence is reached at $n \approx 1000$, which greatly
exceeds the $n=1$ criterion that has frequently been used to indicate the
threshold \cite{bjork_definition_1994,chow_emission_2014,
kreinberg_emission_2017}. To emphasize the contrasting behavior to the
few-emitter nanolaser, the results of Fig.~\ref{fig:Fig2} are shown as dashed
lines. Here, coherence is reached at significantly lower mean photon numbers
($n \approx 10$ for $\beta \rightarrow 1$), and one observes a clear reduction
in the photon number necessary to reach coherence by increasing $\beta$. The
consequences of this finding are highly relevant for the design of future
nanolasers that aim to exploit the unique properties of two-dimensional
TMD and QW gain materials: Here, a reduction in energy consumption requires improving the cavity $Q$ instead of minimizing radiative losses.

The only available $g^{(2)}(0)$ measurements performed in the parameter regime we associate with nanolasers with extended gain media are published in Ref.~[\citenum{jagsch_quantum_2018}]. Upon close inspection, the results shown there indicate that the coherence threshold is also offset to the intensity threshold thereby confirming our prediction. At the same time, the results of our discrete-emitter model are in quantitative agreement with the semiconductor laser model used in the reference, which indeed supports the sufficiency of our approach. For the parameter regime of few-QD nanolasers, recent results have revealed mean photon numbers of $n \approx 100$ at the threshold without an offset between the intensity and coherence thresholds \cite{kreinberg_emission_2017}, which is exactly what is observed in Figs.~\ref{fig:Fig2} and \ref{fig:Fig5} for a similar value of $\beta$.

The origin of the delayed formation of coherence and the insensitivity of the coherence threshold to the $\beta$ factor can be understood  from the contributions to the photon emission as expressed in Eq.~\eqref{RE2} assuming that coherence arises when stimulated emission dominates. Both stimulated emission $\propto n(2f-1)$ and spontaneous
emission $\propto f$ scale with the amount of gain material $N$. On the other
hand, in the lasing regime the carrier population is close to $f_L$ and
the inversion $2f-1$ is then approximately given by $2f_L-1=\kappa/(2RN)$ and,
therefore, decreases with the number of emitters like $1/N$. Corrections to this
approximation are provided by the right-hand side of Eq.~\eqref{anticrossing} in the Appendix,
which is again of the order $1/N$. This is the reason why establishing coherence
requires large photon numbers when
$N$ is large, and much lower $n$ values in the few-emitter case, as demonstrated in Fig.~\ref{fig:Fig5}. Technically,
this behavior is also present in conventional lasers (such as gas and
edge-emitting lasers) but is rendered irrelevant and goes unnoticed as the
transition to coherent emission occurs sharply on a very narrow interval in such
devices due to low $\beta$ factors.

More insight is obtained from the analytical expression for the photon autocorrelation function. From Eq.~\eqref{g2_of_n} it is clear that as $n$ becomes large, the first terms both in the numerator and the denominator become dominant and $g^{(2)}(0)$
approaches unity. Of course, the precise onset of coherence depends also on all
parameters, but the dependence on $\beta$ appears through $\Gamma$ and $R$,
which contain the sum of the loss rates $\gamma + \kappa$. Operating at higher cavity losses, the extended active medium system becomes insensitive to the rate of radiative loss. Two important conclusions are drawn from this observations. On the one hand,
the benefits of high-$\beta$ nanolasers in terms of $n=1$ at threshold and low
threshold pump powers due to increasing $\beta$ rely on being close to the true
few-emitter limit of nanolasing. On the other hand, in systems with many
emitters or a large number of charge carriers in extended media, the influence
of cavity losses on $P_\mathrm{coh}$ dominates that of radiative losses in
device regimes relevant to high-$\beta$ room-temperature lasing with 2D-gain
media. Increasing $\beta$ in those regimes becomes less important than aiming
for improved cavity design with higher quality factors $Q$.

\section{Conclusion}
The prospect of strongly reducing the excitation power required to reach the
threshold by minimizing radiative losses (i.e.~$\beta\rightarrow 1$) is a key
motivation in the development of nanolasers. By going beyond rate equations, we demonstrate that while this is indeed possible in operational regimes found in high-$Q$ nanolasers with few emitters, novel devices based on extended gain media behave differently: There, coherence is established irrespective of the value of $\beta$. Having defined closed expressions for the laser intensity and coherence thresholds (the first indicating the intensity jump, the second the transition to coherent emission), we are able to identify that for the latter, the pump rate at which coherent emission is achieved is rather determined by the cavity losses and can only be improved by increasing $Q$ rather than improving $\beta$. Our results stress the importance of statistical properties of the light emission, given by $g^{(2)}(0)$, and we provide an analytic expression that can be used in conjunction with rate-equation theories to access $g^{(2)}(0)$ to very good approximation. In the light of strong interest in novel ways to design nanolasers by means of new gain media and device geometries, our findings point to possible limitations of future nanolasers based on quantum-well and transition-metal dichalcogenide gain materials. More insight will be obtained once results on the statistical properties of such devices become available.

\begin{acknowledgments}
The authors gratefully acknowledge funding by the DFG (Deutsche 
Forschungsgemeinschaft) and the DFG graduate school ``Quantum Mechanical 
Material Modelling QM$^3$''. P.G.~would like to thank the University of Bremen for 
their hospitality and the DFG for funding the visit and acknowledges 
financial support from CNCS-UEFISCDI Grant No.~PN-III-P4-ID-PCE-2016-0221.
\end{acknowledgments}

\appendix

\section{Derivation of the rate equations}
\label{sec:rate}
The system consists of $N$ identical two-level emitters in resonance with a
cavity mode. In the rotating frame the Hamiltonian reads ($\hbar=1$ throughout)
\begin{equation}
H = \sum_i [g \, b^{\dagger}v^{\dagger}_i c_i + g^* b \, c^{\dagger}_i v_i]
\end{equation}
with $b^{\dagger}, b$ operators for the photon mode, $c^{\dagger}_i, c_i$ for
the upper (conduction-band) emitter states and $v^{\dagger}_i, v_i$ for the
lower (valence-band) ones. We assume a single carrier per emitter, so that the
$c,v$ language is equivalent to the pseudospin formalism by $\sigma_i =
v^{\dagger}_i c_i$ and $\sigma^{\dagger}_i = c^{\dagger}_i v_i$.

The equation of motion (EOM) for the expectation value of a given operator $A$
contains, beside the coherent, von Neumann part, the dissipation expressed in
the Lindblad form
\begin{align}
\frac{d}{dt} \braket{A} = &-i \braket{[A,H]} \nonumber \\
     & + \sum_{\alpha} \frac{\lambda_\alpha}{2}
     \braket{[L^{\dagger}_\alpha,A] L_\alpha + L^{\dagger}_\alpha [A,
      L_\alpha]} \, ,
      \label{vNL}
\end{align}
where $\lambda_\alpha$ is the rate associated to the scattering process defined
by the operator $L_\alpha$. Three such processes are considered: the
radiative loss in each emitter with $\lambda_\alpha$ denoted by $\gamma$ and
$L_\alpha=v^{\dagger}_i c_i$, the cavity losses with the rate $\kappa$ and
operator $b$ and the pumping simulated as an upscattering process, $L_\alpha =
c^{\dagger}_i v_i$, with the rate $P$.

We apply Eq.~\eqref{vNL} to calculate the photon number
$n=\braket{b^{\dagger}b}$ and the upper level population
$f=\braket{c^{\dagger}_i c_i}$. Note that the latter does not depend on the
emitter index $i$. One obtains
\begin{align}
\frac{d}{dt}n = & - \kappa n + 2 N \, \text{Re} \psi \, ,\\
\frac{d}{dt}f = & -\gamma f + P(1-f) -2\, \text{Re} \psi \, .
\end{align}
Both are expressed in terms of the real part of the photon-assisted
polarization $\psi=-ig\braket{b^{\dagger}v^{\dagger}_i c_i}$. The EOM
for $\psi$ brings in higher-order expectation values, such as
$\braket{b^{\dagger}b c^{\dagger}_i c_i}$. Such terms are treated in a
mean-field approximation, which amounts to the truncation at the doublet level
in the cluster expansion formalism  \cite{kira_semiconductor_2011,florian_equation--motion_2013}
\begin{equation}
\braket{b^{\dagger}b c^{\dagger}_i c_i} \approx
\braket{b^{\dagger}b} \braket {c^{\dagger}_i c_i} = n\, f \,.
\end{equation}
The evolution of $\psi$ also generates terms involving polarization operators
$v^{\dagger}c \, , c^{\dagger}v$ from pairs of different emitters. They are
discarded on
reasons discussed in the next section.  As a result one obtains
\begin{equation}
\frac{d}{dt}\psi = -\frac{\Gamma}{2} \psi +|g|^2 n\, (2f-1) + |g|^2 f \, ,
\end{equation}
where $\Gamma=P+\gamma+\kappa$. This also shows that $\psi$ is, in fact, a real
quantity. The steady-state solution is identified by setting the
time derivatives to zero. By eliminating $\psi$ in the resulting equations
one obtains for the populations
\begin{subequations}
\begin{align}
& R \,n (2f-1) + R f =  -\gamma f + P(1-f) \, ,\label{rateq_f} \\
& N [R\,n (2f-1) + R f] = \, \kappa n \, ,
\label{rateq_n}
\end{align}
\end{subequations}
with the rate of spontaneous emission $R$ given by
\begin{equation}
R = \frac{4 |g|^2}{P + \gamma + \kappa} \, .
\end{equation}

The equations above are balance conditions. Their left-hand side contains
the net photon generation rate, made up of spontaneous and stimulated emission,
as well as absorption. In Eq.~\eqref{rateq_f} this is balanced against the rate
of the emitter excitation, taking into account both the pumping and the
radiative losses, while in Eq.~\eqref{rateq_n} the photons generated by all
emitters compensate the cavity losses.

Combining these equation one obtains
\begin{equation}
f_0-f=\frac{\kappa n}{N(P+\gamma)}~,
\qquad \mathrm{with} \qquad f_0=\frac{P}{P+\gamma}.
\label{f_0}
\end{equation}
It is obvious that $f_0$ is the steady-state upper-level population in the
absence of the Jaynes-Cummings interaction ($R=0$).

Eq.\ \eqref{rateq_n} can be rewritten as
\begin{equation}
f_L-f=\frac{f}{2n}~,
 \qquad \mathrm{with}\qquad
 f_L=\frac{\kappa}{2RN}+\frac{1}{2},
\label{f_L}
\end{equation}
where $f_L$ is the population for which the gain (stimulated emission
minus absorption) exactly compensates the cavity losses.
As seen from the above
equations, one has  $f\leqslant f_0, f_L $ i.e.~both $f_0$ and $f_L$ are upper
bounds for the true steady-state solution.
Multiplying Eqs. \eqref{f_0} and \eqref{f_L} one obtains a quadratic equation
for $f$
\begin{equation}
(f_L-f)(f_0-f)=\frac{\kappa f}{2(P+\gamma)N},  \label{anticrossing}
\end{equation}
whose lower solution is the physical one. In turn, the photon population is
given by
\begin{equation}
n=\frac{N(P+\gamma)}{\kappa}\, (f_0-f).  \label{ph_number}
\end{equation}
This solves the problem of steady-state populations for the carriers and for the photons at the rate-equation level.

In the limit $N \to \infty$ the rhs of Eq.\ \eqref{anticrossing} vanishes and,
as the pump increases, the solution changes abruptly from $f=f_0$ to $f=f_L$,
always following the smaller. The $f=f_0$ case corresponds to pumping spent
exclusively for exciting the carrier subsystem. This takes place until $f$
reaches the value $f_L$ with sufficient inversion to compensate the photon
losses. From then on massive photon generation takes place, as seen in Eq.\
\eqref{ph_number}, which shows the photon density $\eta=n/N$ changing sharply
from $\eta = 0$ to $\eta \sim f_0-f > 0$. This is the lasing regime (hence the
index $L$), in which the photon number becomes macroscopic, i.e.~scales like
$N$. It becomes clear that in the limit $N \to \infty$ one obtains a sharp
crossover from a ``normal'' regime to lasing. The threshold is defined by the
pump value $P_{\mathrm{th}}$, which obeys the Eq.~\eqref{Pthr}, expressing the crossing condition $f_0=f_L$.

For finite values of $N$, the transition is no longer abrupt. The rhs of
Eq.~\eqref{anticrossing} being nonzero, a smooth change (anticrossing)
from $f \approx f_0$ to $f\approx f_L$ takes place. The lasing transition
becomes gradual, without a well-defined threshold. Now a transition interval
around $P_{\mathrm{th}}$ still separates two contrasting behaviors: Above it the photon number grows with system size $N$, while below it their number stays finite.

\section{Beyond rate equations: accessing $g^{(2)}(0)$}
The second-order autocorrelation function requires the cluster expansion at
least up to the quadruplet level. This introduces a set of new expectation
values besides the populations $n=\braket{b^{\dagger}b}$ and
$f=\braket{c_i^{\dagger}c_i}$. The following notations are used below:
\begin{equation}
c_m=\braket{b^{\dagger m}b^m c_i^{\dagger}c_i} \, , \,
v_m=\braket{b^{\dagger m}b^m v_i^{\dagger}v_i} \, , \,m = 0,1,2 \,
\end{equation}
\begin{equation}
p_m=\braket{b^{\dagger m}b^m} \, , \,
\psi_m=-i g\braket{b^{\dagger m} b^{m-1}v_i^{\dagger}c_i}\, , \, m= 1,2
\end{equation}
With a single carrier in each emitter, these averages are not
independent since, obviously $c_m+v_m = p_m$, but we keep the notation for
convenience. Also, $p_1$ is the average photon number $n$, $c_0=f$ and $\psi_1$
is the same as $\psi$ of the previous section. The index
$i$ is spurious here, since all emitters are identical.

 \begin{figure}[t]
   \centering
    \includegraphics[width=0.5\textwidth]{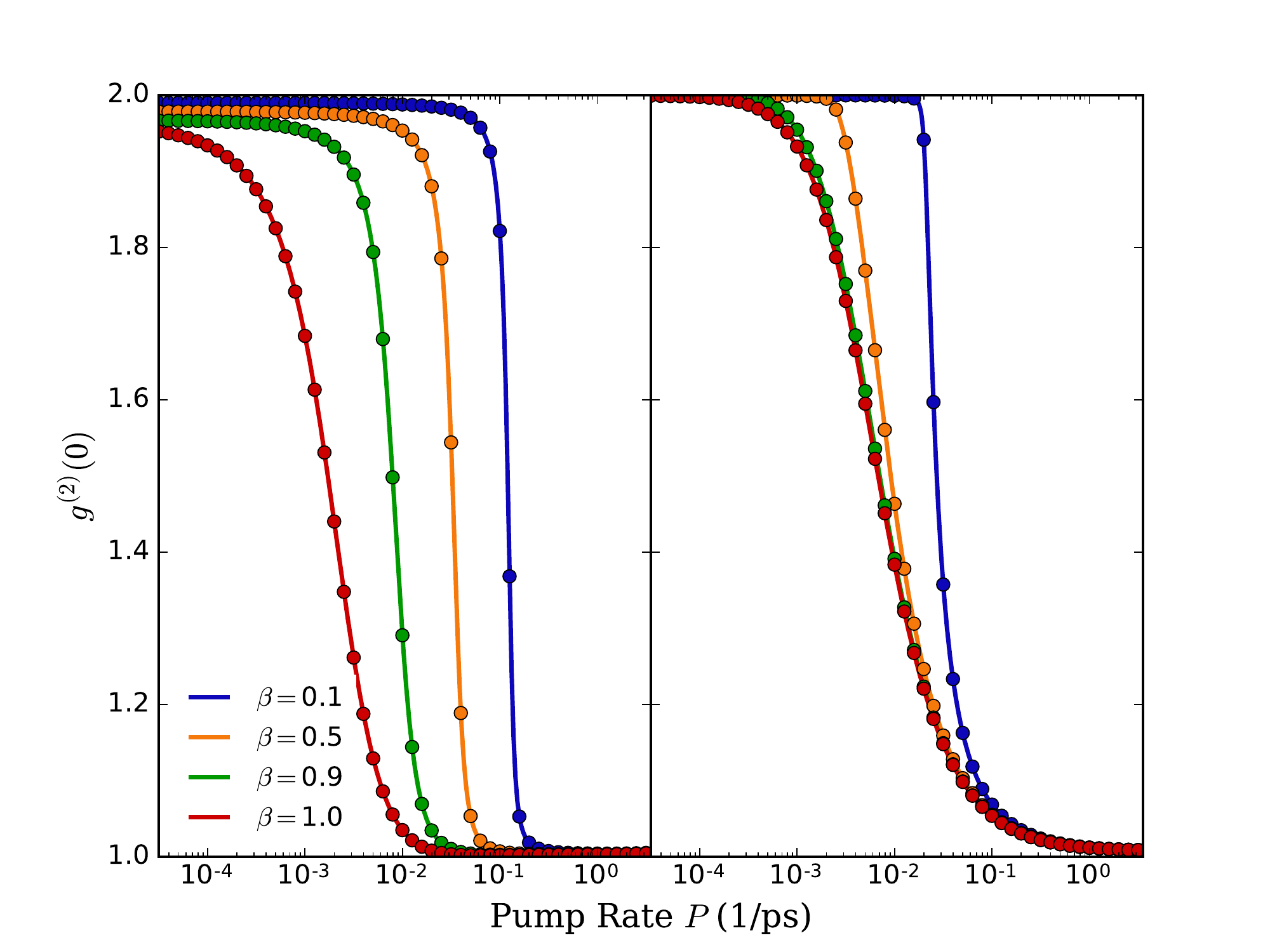}
    \caption{Photon autocorrelation function $g^{(2)}(0)$ obtained by using
    full quadruplet (solid) and rate-equation (circles) populations.
    Both the few- (left) and many-emitter (right) cases are represented.
      \label{fig:SM_Fig1}}
 \end{figure}
 
Quadruplet averages generated by equations of motion for doublet
quantities are not factorized any more. As an example, consider the
case of the one-photon-assisted polarization $\psi_1$, for a generic emitter $i$
\begin{equation}
\frac{d}{dt} \psi_1= |g|^2 (c_1-v_1 +c_0) -\frac{1}{2}\Gamma \psi_1
     + |g|^2 \sum_{j \neq i} \braket{c^{\dagger}_j v_j v^{\dagger}_i c_i} \, .
\label{psi1}
\end{equation}
In contrast to the rate equation, where one approximates
$c_1 = p_1 c_0$, and $v_1=p_1 v_0 $, here we write new EOM for them. Having an
additional $b^{\dagger}b$ factor in its definition, the evolution of
$c_1$ generates the two-photon-assisted polarization $\psi_2$. The same
holds for $p_2$, which is the central quantity of the two-photon autocorrelator
$g^{(2)}(0)$. The equation for $\psi_2$ reads
\begin{align}
\frac{d}{dt} \psi_2=& |g|^2 (c_2-v_2 +2 c_1) -\frac{1}{2}\Gamma' \psi_2
     \nonumber \\
     + & 2 |g|^2 \sum_{j \neq i} \braket{b^{\dagger}b c^{\dagger}_j
       v_j v^{\dagger}_i c_i}
     \nonumber  \\
     - & g^2 \sum_{j \neq i} \braket{b^{\dagger}b^{\dagger} v^{\dagger}_j
       c_j v^{\dagger}_i c_i} \, ,
\label{psi2}
\end{align}
where $\Gamma' = P+\gamma+3 \kappa$. Quantities such as $c_2$ and $v_2$ are
cluster expanded in terms of correlators as  \cite{kira_semiconductor_2011,florian_equation--motion_2013}
\begin{equation}
c_2 = \delta c_2 +\delta p_2 \,c_0 +4 \delta c_1 \,p_1 +2 p_1^2\,c_0 \, ,
\end{equation}
and similarly for $v_2$. As a sixtuplet correlator, $\delta c_2$ is neglected,
and using $\delta p_2 = p_2-2\, p_1^2$, $\delta c_1 =c_1 -p_1\, c_0$, one
obtains the expression of $c_2$ in its truncation at the quadruplet level as
\begin{equation}
c_2 = p_2\,c_0 + 4 p_1 c_1 - 4 p_1^2\, c_0 \, .
\end{equation}
The summations appearing in  both Eq.~\eqref{psi1} and \eqref{psi2} describe
correlations of polarizations and photon-assisted polarization, respectively,
between pairs of emitters. Bearing in mind that quantities on different sites
are less correlated than same-site ones, we adopt the procedure of factorizing
into averages on separate sites. For example,
\begin{equation}
\braket{c^{\dagger}_j v_j v^{\dagger}_i c_i} \approx
\braket{c^{\dagger}_j v_j}\braket{v^{\dagger}_i c_i} \, .
\end{equation}
This expresses the sum in Eq.~\eqref{psi1} in terms of on-site polarizations.
But these are anomalous averages, not driven by the system excitation, and
as such are vanishing quantities. Therefore this sum is taken as zero. This is
not the case of the two sums in Eq.~\eqref{psi2}, where by the same procedure
we get
$2(N-1)\psi_1 \psi_1^*$ and $(N-1)\psi_1^2$, respectively. Since the equations
obeyed by $\psi_1$ show that this is a real quantity, the two expressions are
equal up to the prefactor, and a partial cancelation takes place, leaving the
net result as $(N-1)\psi_1^2$. Let us emphasize that such terms, which express
correlations between two emitters mediated by the photon field are usually
neglected when one is not interested in superradiance effects  \cite{leymann_sub-_2015}. In
the present approach they are included at a mean-field level.

The other EOM do not generate higher-order averages. As a result of the
procedure described above, one is left with a closed system of EOM for $c_0,
c_1, p_1, p_2, \psi_1$ and $\psi_2$. In the steady state the time derivative
vanishes and the system becomes algebraic. From this, one obtains an expression
of $p_2$ and thus of $g^{(2)}(0)$ in terms of $c_0=f$ and $p_1=n$ only, by
eliminating the other unknowns. The result is expressed in Eq.~(6) of the main
text. Further simplification arises by eliminating the upper-level population $f$
in favor of the photon number $n$ and inserting the rate-equation value of the latter,
as described in the derivation of Eq.~(8) of the main text. The approximation is
very accurate, as shown in Fig.~\ref{fig:SM_Fig1}, which compares it to the result of using full quadruplet-computed populations in Eq.~(6).

\bibliography{Lohof_et_al_submission}

\end{document}